\documentclass[twocolumn,tighten]{aastex7}
\usepackage{graphicx}
\usepackage{amsmath}
\usepackage{scalerel}
\usepackage{apjfonts}
\usepackage{tikz}
\usetikzlibrary{svg.path}

\definecolor{orcidlogocol}{HTML}{A6CE39}
\tikzset{
  orcidlogo/.pic={
    \fill[orcidlogocol] svg{M256,128c0,70.7-57.3,128-128,128C57.3,256,0,198.7,0,128C0,57.3,57.3,0,128,0C198.7,0,256,57.3,256,128z};
    \fill[white] svg{M86.3,186.2H70.9V79.1h15.4v48.4V186.2z}
                 svg{M108.9,79.1h41.6c39.6,0,57,28.3,57,53.6c0,27.5-21.5,53.6-56.8,53.6h-41.8V79.1z M124.3,172.4h24.5c34.9,0,42.9-26.5,42.9-39.7c0-21.5-13.7-39.7-43.7-39.7h-23.7V172.4z}
                 svg{M88.7,56.8c0,5.5-4.5,10.1-10.1,10.1c-5.6,0-10.1-4.6-10.1-10.1c0-5.6,4.5-10.1,10.1-10.1C84.2,46.7,88.7,51.3,88.7,56.8z};
  }
}

\newcommand\orcidicon[1]{\href{https://orcid.org/#1}{\mbox{\scalerel*{
\begin{tikzpicture}[yscale=-1,transform shape]
\pic{orcidlogo};
\end{tikzpicture}
}{|}}}}

\usepackage{comment}
\usepackage{array,multirow,color,tablefootnote}
\usepackage{xcolor}

\newcommand{\um}{$\mu$m}

\newcommand{\slopeunit}{$\%/100\,\mathrm{nm}$}

\begin{document}

\title{Evidence of Possible Spectral Variability in the Patroclus--Menoetius Binary System}

%===============================================================================

\author[orcid=0000-0001-9665-8429]{Ian~Wong}
\affiliation{Space Telescope Science Institute, Baltimore, MD, USA}
\email{iwong@stsci.edu}

\author[orcid=0000-0002-8296-6540]{William~M.~Grundy}
\affiliation{Lowell Observatory, Flagstaff, AZ, USA}
\affiliation{Northern Arizona University, Flagstaff, AZ, USA}
\email{w.grundy@lowell.edu}

\author[orcid=0000-0001-9265-9475]{Joshua~P.~Emery}
\affiliation{Northern Arizona University, Flagstaff, AZ, USA}
\email{joshua.emery@nau.edu}

\author[orcid=0000-0002-9995-7341]{Richard~P.~Binzel}
\affiliation{Department of Earth, Atmospheric and Planetary Sciences, Massachusetts Institute of Technology, Cambridge, MA, USA}
\email{rpb@mit.edu}

\author[orcid=0000-0002-1700-5364]{Oriel~A.~Humes}
\affiliation{Institut für Geophysik und Extraterrestrische Physik, Technische Universität Braunschweig, Braunschweig, Germany}
\email{oriel.humes@tu-braunschweig.de}

\author[orcid=0000-0003-2548-3291]{Simone~Marchi}
\affiliation{Southwest Research Institute, Boulder, CO, USA}
\email{marchi@boulder.swri.edu}

\author[orcid=0000-0002-4725-4775]{Pippa~M.~Molyneux}
\affiliation{Southwest Research Institute, San Antonio, TX, USA}
\email{pmolyneux@swri.edu}

\author[orcid=0000-0002-6013-9384]{Keith~S.~Noll}
\affiliation{NASA/Goddard Space Flight Center, Greenbelt, MD, USA}
\email{keith.s.noll@nasa.gov}

%===============================================================================
\begin{abstract}
We present new visible-wavelength spectroscopic observations of the Patroclus--Menoetius binary system in the Jupiter Trojan population. Motivated by previously published spectra from different instruments that showed evidence of significant longitudinal variability, we obtained two spectra spanning 440--680\,nm at near-opposite rotational phases with the Gemini Multi-Object Spectrograph on the Gemini South telescope during the late 2024 apparition. The same solar analog was used for both observations to remove one source of inconsistency. We measured spectral slopes of $2.51\%\pm0.05\%$/100\,nm and $8.13\%\pm0.05$\%/100\,nm at the two different rotational phases. The first of these measurements was serendipitously obtained during an occultation of Menoetius by Patroclus. Although the statistical significance of the spectral slope discrepancy persists even after considering possible systematic errors stemming from differences in slit position angles and air masses between the asteroid and solar analog exposures, we consider this report of variability to be tentative. We briefly explore several scenarios that could explain the measured spectral slope variability. Additional follow-up observations are necessary to definitively confirm and characterize any inhomogeneities across the surface, which will have major implications for the 2033 flyby of Patroclus--Menoetius by the Lucy spacecraft.
\end{abstract}


%===============================================================================
\section{Introduction}
\label{sec:intro}
%===============================================================================

In 2021 October, the Lucy spacecraft \citep{levison2021,olkin2021} set off on its decade-long journey to explore the Jupiter Trojans. Lucy will reach its final flyby target, the Patroclus--Menoetius binary system, in 2033. Patroclus--Menoetius (hereafter, PM binary) is the second-largest Jupiter Trojan, with volume-equivalent spherical component diameters of 113 and 106\,km, respectively \citep{buie2015}, and one of only five confirmed multiple systems in this population \citep{merline2001,marchis2006,noll2016,noll2020,buie2022}.

Observations of mutual events between the primary and secondary and relative astrometry of the resolved binary refined the sizes and orbit of the two components \citep[e.g.,][]{buie2015,grundy2018,brozovic2024}, placed constraints on the level of small-scale surface heterogeneity in visible color \citep{wong2019}, and detected a large crater-like depression on the secondary \citep{pinillaalonso2022}. These studies also provided a low bulk density measurement of $0.81\pm0.16$\,g\,cm$^{-3}$ \citep{berthier2020}---indicating a highly porous water-ice-rich interior---and low thermal inertia values, consistent with fine-grained regolith \citep{mueller2010}. Ground-based spectroscopy spanning the visible and near-infrared up to 2.5\,\um\ has produced a featureless reflectance spectrum \citep{luu1994,emery2003,yang2007,emery2011,sharkey2019}, with a visible color and continuum shape that places the PM binary within the less-red of the two attested Trojan color subpopulations \citep[e.g.,][]{szabo2007,roig2008,emery2011,wong2014}. Spectroscopic observations with JWST targeting longer wavelengths revealed a broad absorption feature spanning 2.7--3.5\,\um\ \citep{wong2024}---attributable to OH-bearing species and possibly an additional minor contribution from aliphatic organics---robustly confirming a previous finding from ground-based $L$-band measurements \citep{brown2016}.

A recent study uncovered evidence of possible large-scale longitudinal variability in the PM binary system. \citet{humes2022} presented two spectroscopic datasets obtained with two different telescopes: (1) near-ultraviolet and visible (0.2--1.0\,\um) slitless spectroscopy with the Wide Field Camera 3 on the Hubble Space Telescope (HST) and (2) visible-wavelength measurements spanning 0.35--1.0\,\um\ from the DeVeny spectrograph on the Lowell Discovery Telescope (LDT). Although both spectra are featureless at visible wavelengths, they exhibit drastically different continuum shapes. While the LDT spectrum shows a monotonically increasing spectral slope of 8.7\slopeunit\ that is consistent with the color of typical P-type asteroids within the Tholen classification, the HST spectrum is significantly flatter overall, with a measured spectral slope of 3.5\slopeunit. Moreover, the HST spectrum exhibits a peculiar inflection point at $\sim$0.6\,\um, with the spectral slope decreasing to near zero at longer wavelengths. While the subobserver latitudes of the HST and LDT observations were similar ($+12^{\circ}$ and $-7^{\circ}$, respectively), the subobserver longitudes differed by 105$^{\circ}$, raising the possibility of large-scale variations in spectral properties across the PM binary system. However, due to differences in instrumentation, observing epochs, data processing workflows, and possible uncorrected systematics between the two observations, caution is warranted when interpreting these results. Follow-up observations using a consistent observational setup are needed to further probe for longitudinal variability on the PM binary. To this end, we obtained a pair of ground-based visible spectra of the PM binary targeting roughly opposite rotational phases using the Gemini Multi-Object Spectrograph (GMOS; \citealt{hook2004,gimeno2016}) on the Gemini South telescope.

%===============================================================================
\section{Observations and Data Analysis}
\label{sec:obs}
%===============================================================================

\begin{figure}[t!]
\centering
    \includegraphics[width=0.85\columnwidth]{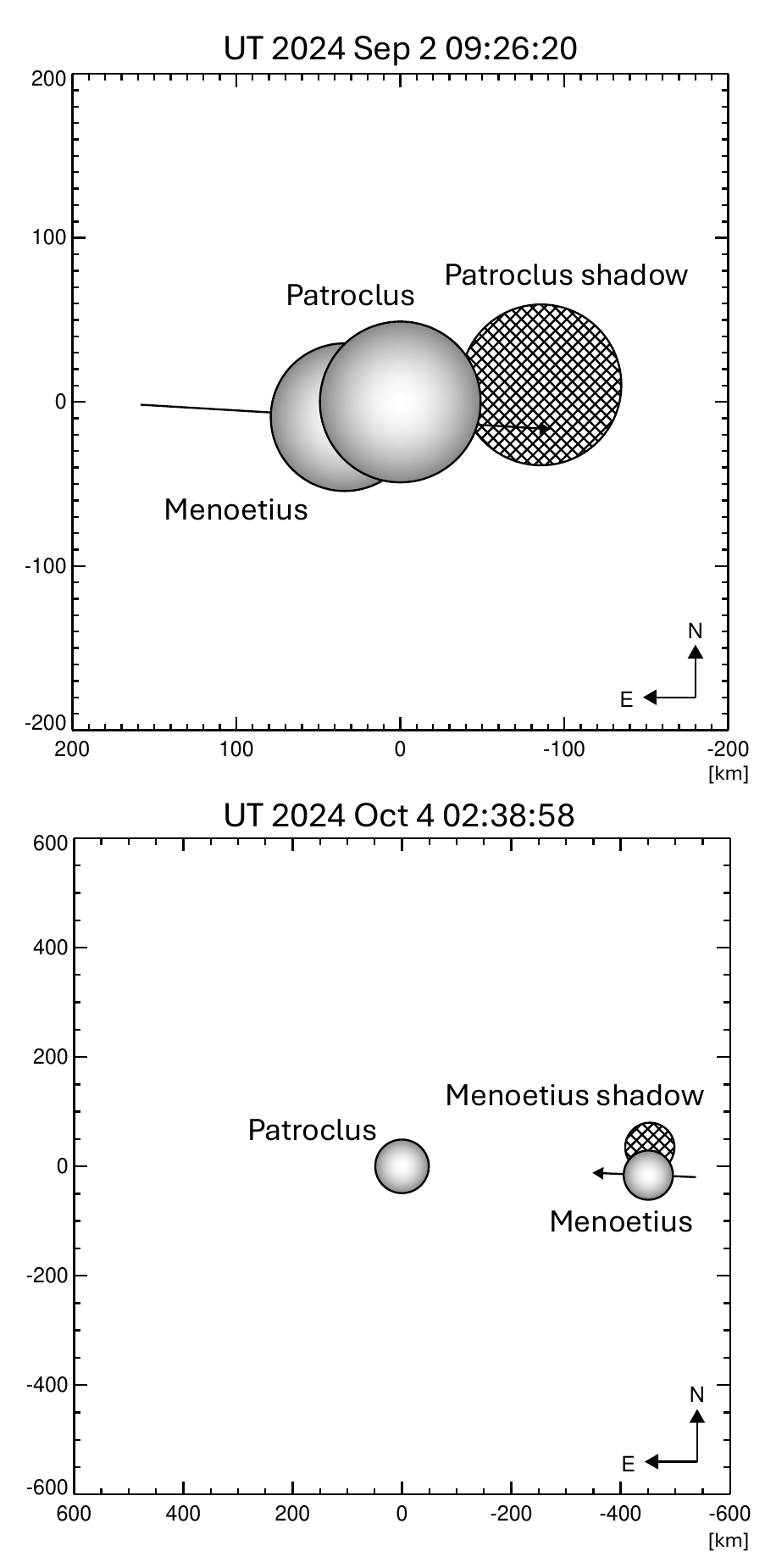}
    \caption{Viewing geometries at the start times of the GMOS observations, centered in the stationary frame of Patroclus. The sky-projected disks of Patroclus and Menoetius are approximated by circles with radii of 49 and 45\,km, respectively. The hatched circles indicate the shadow of Patroclus or Menoetius. The direction of Menoetius's instantaneous relative motion is indicated by the black arrows. The ``N'' and ``E'' directions denote increasing decl. and R.A., respectively. The rotational poles of the two tidally locked components point downward in these plots.}
    \label{fig:geom}
    \vspace{-0.2cm}
\end{figure}

\begin{deluxetable*}{lccccccccccccccccccc}[t!]
\setlength{\tabcolsep}{1.5pt}
\tablewidth{0pc}
\renewcommand{\arraystretch}{0.9}
\tabletypesize{\small}
\tablecaption{
    GMOS Observation Details and Viewing Geometries
\label{tab:obs}}
\tablehead{
    Target & $\qquad$ & UT Start Time & $\qquad$ & Air Mass & $\qquad$ & Slit Offset & $\qquad$ & $\Delta$  & $\qquad$ & $r$ & $\qquad$ & $\alpha$ & $\qquad$ & $\lambda$ & $\qquad$ & $\beta$ & $\qquad$ & $d$ \\
    & & & & & & (deg) & & (au) & & (au) & & (deg) & & (deg) & & (deg) & & (km)
}
\startdata
PM binary & & 2024 Sep 2 09:26:20 & & 1.27 & & 66 & & 3.61 & & 4.48 & & 7.2 & & 183 & & 0.9 & & 35 \\
HD~6461 & & 2024 Sep 2 09:48:06 & & 1.35 & & 0 \\
\hline
PM binary & & 2024 Oct 4 02:38:58 & & 1.16 & & 48 & & 3.53 & & 4.49 & & 4.2 & & 41 & & 0.2 & & 450 \\
HD~6461 & & 2024 Oct 4 03:03:46 & & 1.17 & & 48 \\
\enddata
\footnotesize{
\vspace{6pt}
\noindent \textbf{Note.} The slit offset is the difference between the slit position angle and the local parallactic angle at the start of the observation. $\Delta$, $r$, and $\alpha$ are the geocentric distance, heliocentric distance, and phase angle of the PM binary's barycenter at the start of the observation. $\lambda$ and $\beta$ denote the subobserver longitude and latitude as computed using the mutual orbit solution in \citet{pinillaalonso2022}, with $\lambda = 0^{\circ}$ and $\beta = 0^{\circ}$ corresponding to the sub-Menoetius point on Patroclus. $d$ is the sky-projected separation of the two components at the start of the observation.}
\vspace{-22pt}
\end{deluxetable*}

\subsection{Observing Strategy}\label{subsec:setup}
Queue-scheduled observations of the PM binary with GMOS were carried out on UT 2024~September~2 and October~4 during its late 2024 apparition. To maximize the self-consistency between our two observing epochs, we utilized the same exposure setup, solar analog, and data reduction framework for both visits. Each visit consisted of a single 300\,s exposure of the PM binary, with the telescope tracking its nonsidereal motion. All observations utilized the B480-G5327 grating and the $2\overset{''}{.}0$ wide, $5\overset{'}{.}5$ long single slit. The observation of the science target was immediately followed by an analogous 3\,s exposure of the nearby G-type star HD~6461. This star was chosen from a list of bright ($V < 10$\,mag) solar-type stars to minimize the distance between the star and the PM binary across both visits. HD~6461 is a red horizontal branch star, with an effective temperature of $\sim$5100\,K that places it well below the instability strip where RR Lyrae stars lie \citep[e.g.,][]{behr2003}; it is therefore not expected to display significant variability. 

Based on the mutual orbit ephemeris and rotational pole solution in \citet{pinillaalonso2022}, the two observations targeted subobserver longitudes of 183$^{\circ}$ and 41$^{\circ}$, respectively; the subobserver latitudes for the two visits were nearly identical---$0\overset{\circ}{.}9$ and $0\overset{\circ}{.}2$, respectively. The first visit on 2024~September~2 serendipitously coincided with a superior occultation event---when Menoetius's sky-projected disk passed behind Patroclus. Figure~\ref{fig:geom} presents the viewing geometries of the PM binary at the start of the two visits. The components are approximated by circles with radii corresponding to the smallest (i.e., polar) dimension of the corresponding triaxial shape models from \citet{buie2015}. The longitude--latitude array is defined with respect to Patroclus's surface, with the zero-point set at the sub-Menoetius point. The grid orientation uses the right-handed convention relative to the rotational pole, which in this case is retrograde (i.e., pointing downward in the figure). The sky-projected separations of the PM binary were 35 and 450\,km for the two visits, respectively, corresponding to on-sky distances of $0\overset{''}{.}013$ and $0\overset{''}{.}176$. It follows that the binary was not resolved during either observation. Table~\ref{tab:obs} lists the full observational details and viewing geometries for the two GMOS visits.

The adjustable central wavelength was set to 500\,nm, providing near-continuous coverage from 360 to 690\,nm, dispersed across three Hamamatsu CCDs at $2\times$ spatial and $4\times$ spectral binning. Observing conditions were nominal, with no cloud cover and good image quality (designated in the file headers with the code IQ70, i.e., seeing better than $\sim$$0\overset{''}{.}8$). A pair of internal flat field and argon arc lamp exposures was acquired immediately preceding the observations of the PM binary and the solar analog. A set of nightly bias measurements was provided by the observatory.

One complication in the operation of GMOS is guiding. The on-instrument wavefront sensor consists of a small movable lenslet array that can patrol a $\sim$15\,arcmin$^{2}$ area situated in one quadrant of the GMOS science field. If no suitable guide star is located within the patrol region, the instrument alignment angle must be rotated. As a result, although we requested that the slit angle be aligned with the parallactic angle for all exposures, this was only possible for the observation of the solar analog on the first night. To mitigate the effect of differential atmospheric refraction on our spectra, we selected the relatively wide $2\overset{''}{.}0$ slit. We explore the possible impacts of slit alignment in Section~\ref{subsec:systematics}.

\begin{figure*}[t!]
\centering
    \includegraphics[width=0.8\textwidth]{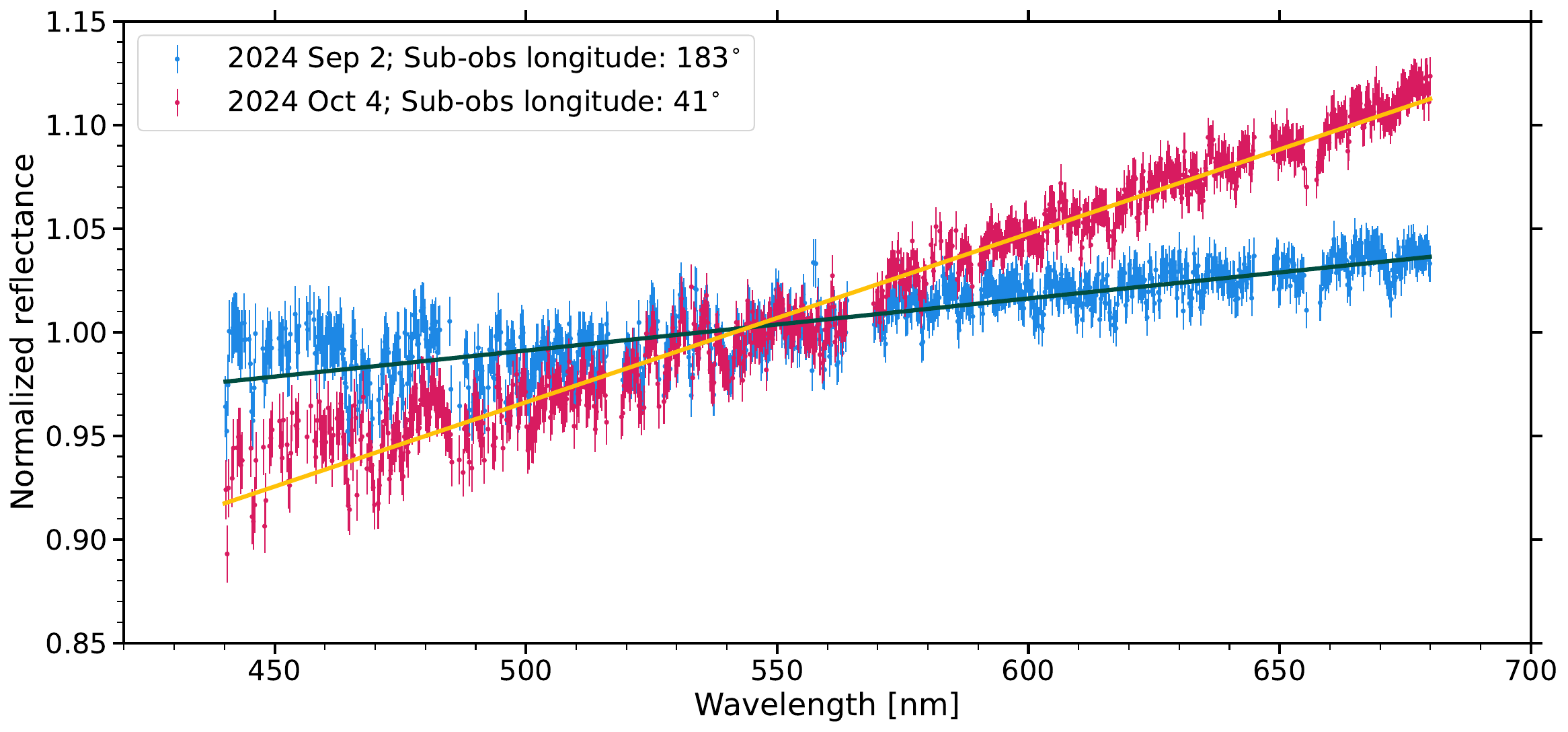}
    \caption{Reflectance spectra of the PM binary from the two visits (blue = 2024~September~2, magenta = 2024~October~4). The best-fit spectral slopes are overplotted in green and yellow, respectively. Both spectra are normalized to unity at 550\,nm. The uncertainties are inflated so that the reduced $\chi^{2}$ statistic of the linear fit is 1.}
    \label{fig:reflectance}
\end{figure*}

\subsection{Data Reduction}

We processed the data and extracted the spectra using the \texttt{DRAGONS} data reduction tool \citep{dragons}. The software was run with default settings for the various steps within the reduction workflow. Several interactive modules are available that allow the user to vet and adjust the wavelength solution, aperture finding, sky background correction, and spectral trace modeling, and we carefully inspected the diagnostic plots to ensure that the spectral extraction and wavelength calibration were optimized. The exclusion threshold used to mask the flux from astronomical sources within the slit was tightened from 0.1 to 0.01 of the maximum flux in the cross-dispersion direction in order to improve the fidelity of the background modeling in the vicinity of the target. The order of the Chebyshev polynomial used for trace fitting was raised from 2 to 3 to better match the shape of the spectral profile across the detectors. After producing bias-corrected, flat-fielded, and wavelength-calibrated combined detector images, we extracted the spectra of the PM binary and HD~6461 using aperture widths of 20, 30, 40, and 50\,px, corresponding to $3\overset{''}{.}2$, $4\overset{''}{.}8$, $6\overset{''}{.}4$, and $8\overset{''}{.}0$, respectively.

\subsection{Reflectance Spectra}

When dividing the extracted spectra of the PM binary and the solar analog from one another, we detected correlated features in the vicinity of the strongest stellar absorption bands, indicating small deviations in the wavelength solution between the arc lamp and on-target exposures. We carried out a cross-correlation analysis to align the stellar features in the spectra. The required shifts in the wavelength grid of the stellar spectra were very small: +0.34 and +0.16\,nm for the two visits, respectively, which are smaller than the average wavelength resolution element of the spectra. For comparison, the magnitude of the Doppler shift due to the radial velocity of the star (7.98\,km/s; \citealt{gaiadr2}) is 0.013\,nm at 500\,nm. Next, we corrected the reflectance spectra for the difference in color between the solar analog ($B-V=0.785$; \citealt{hog2000}) and the Sun; we note that while this correction was needed to rectify the spectral slope measurements, it does not impact the detectability of spectral variations across the PM binary, given that the same star was used for both visits by design.

The spectra extracted using the 20\,px aperture exhibit the smallest scatter, and we present the reflectance spectra derived from those extractions in Figure~\ref{fig:reflectance}. Both spectra are normalized to unity at 550\,nm and are trimmed to the range 440--680\,nm to remove the portions with low throughput and poor signal-to-noise ratio. Some systematic features persist in the spectra near major stellar absorption bands, likely due to differences in the extent of rotational line broadening between HD~6461 and the Sun. 

The continuum slopes of the two spectra are highly discrepant. We measured the spectral slopes using an iterative approach. After an initial linear regression, we removed significant outliers in the residual arrays. Following a second linear fit to the trimmed spectra, we inflated the uncertainties by a constant multiplicative factor of 1.8 to yield a reduced $\chi^{2}$ value of 1. The final fitting was carried out using the Markov Chain Monte Carlo ensemble sampler \texttt{emcee} \citep{emcee}, from which the median and $1\sigma$ uncertainty in the slope were computed. The spectral slopes from the two visits at near-opposite rotational phases are $2.51\%\pm0.05$\slopeunit\ and $8.13\%\pm0.05$\slopeunit, respectively. To search for possible deviations from linearity in the reflectance spectra (e.g., an inflection point similar to the one previously reported from HST observations; \citealt{humes2022}), we experimented with fitting quadratic and cubic polynomials to the spectra but did not obtain statistically significant higher-order coefficients.

\subsection{Possible Sources of Systematic Biases}\label{subsec:systematics}

The most salient potential source of systematic biases in the measured spectral slopes is differential atmospheric refraction. As discussed in Section~\ref{subsec:setup}, the restricted patrol region of the GMOS guider necessitated adjustments to the slit position angle to ensure proper guiding. On 2024~September~2, while the exposure of the solar analog was obtained with the slit aligned with the parallactic angle, the observation of the PM binary was executed with the slit rotated by 66$^{\circ}$ away from the parallactic angle. On 2024~October~4, both the PM binary and HD~6461 were observed with a 48$^{\circ}$ offset between the slit position angle and the parallactic angle.

To explore the impact of differential refraction due to the slit position angle alignments, we followed the slit throughput and atmospheric refraction formalism described in \citet{filippenko1982}.\footnote{These calculations were facilitated by the online tool developed by Santos~Pedraz~Marcos: \url{https://www.caha.es/pedraz/RS/refract_slit.html}.} This analysis focused on the first visit (2024~September~2), when the slit position angle alignments were drastically different for the PM binary and the solar analog. For the second visit, because both exposures were collected with the same relative slit position angle offset, the effects of differential atmospheric refraction are expected to cancel out when deriving the reflectance spectrum. 

We entered the measured atmospheric conditions (ambient temperature, air pressure, and humidity) and the PM binary's air mass at the time of the observation, which are contained in the header of the raw data file. When assuming that the target was centered in the slit at 500\,nm, we found that the predicted differential loss in slit throughput at 650\,nm for a slit misalignment of 66$^{\circ}$ and $0\overset{''}{.}8$ seeing is roughly 0.3\%, corresponding to a negative spectral slope bias of $\sim$0.2\slopeunit. This value is negligible when compared to the difference between the measured slopes from the two visits. We also note that if the PM binary was instead centered in the slit at 650\,nm, the slope bias switches sign, thereby slightly increasing the discrepancy between the spectral slopes obtained for the two opposite hemispheres.

Another potential source of bias is relative air mass. Because we used the same star for both visits, the relative air masses of the asteroid and the star changed across the roughly one month interval between the two visits. While both targets were situated at nearly identical air masses during the second visit, the difference in air mass between HD~6461 and the PM binary was 0.08 during the first visit. The higher relative air mass of the solar analog during the 2024~September~2 observations would have imparted some additional reddening to its spectrum, resulting in a negative bias in the resultant spectral slope measurement from the PM binary's reflectance spectrum. To estimate the expected magnitude of this effect, we used the extinction curve reported for Cerro Tololo \citep{stone1983}, which is located 10\,km to the southeast from the Gemini South telescope at Cerro Pachón; no published extinction data are available for Cerro Pachón. The extinction coefficients at 450 and 650\,nm are 0.22 and 0.10\,mag/air mass, respectively. It follows that an air mass differential of 0.08 would yield a relative spectral slope difference of roughly 0.4\slopeunit.

We also empirically quantified the magnitude of the combined effects of differing air masses, slit position angles, and other observing conditions by directly dividing the stellar spectra obtained from the two visits from one another. A linear fit to the resultant ratio array produced a differential slope measurement of 0.7\slopeunit. This value dwarfs the statistical uncertainty obtained from our spectral slope fitting while also exceeding the estimated slope bias in the first visit due to differential air mass alone. We therefore take 0.7\slopeunit\ as an estimate of the inherent systematic uncertainty of our spectral slope calculations. Even when considering this enhanced uncertainty, the spectral slope measurements obtained for the two opposite hemispheres of the PM binary differ at the $5.7\sigma$ level.

%===============================================================================
\section{Discussion}
\label{sec:disc}
%===============================================================================

Our GMOS observations provide evidence of large-scale variability within the PM binary system. However, although the statistical significance of the spectral slope discrepancy remains high even after considering potential sources of systematic biases between the two observations, we still urge caution when interpreting these results and consider our detection of spectral slope variations to be tentative. We cannot exclude the possibility of unmodeled sources of systematic biases related to the interplay between optical path, slit throughput, and atmospheric refraction, along with other specifics of the observing conditions across the two visits. Follow-up observations, preferably using space-based facilities and/or slitless spectrographs (e.g., integral field units), are needed to definitively confirm and characterize any spectral variability across the PM binary. High-cadence multiband photometry or spectroscopy of upcoming mutual events would be especially valuable for mapping the spectral slope distribution and probing surface inhomogeneities.

For now, we explore two possible scenarios that could produce large-scale variations in the spectral properties of the PM binary. In the first scenario, Patroclus and Menoetius have different visible colors but are individually homogeneous across their surfaces. The coincidental timing of the first visit during a mutual event provides a unique perspective on spectral variations across the system. While the spectral slope measured from the second visit reflects the average disk-weighted color of the two similarly sized components, the shallower slope obtained from the first visit is impacted by the partial removal of Menoetius's contribution from the total system flux due to occultation by Patroclus. In this instance, Patroclus must have a significantly more neutral surface color than Menoetius. Using a disk-area-weighted linear mixture model (see Equation~(4) in \citealt{wong2019}), we computed the eclipsed area of Menoetius during the first visit and converted the measured spectral slope discrepancy to the systematic color difference between Patroclus and Menoetius in this scenario: 34\slopeunit.

This inferred color discrepancy far exceeds the full range of spectral slopes measured from Sloan Digital Sky Survey photometry \citep[e.g.,][]{wong2014}. Meanwhile, time-resolved photometric observations of a previous inferior shadowing event (i.e., when Menoetius's shadow passed over Patroclus) found no statistically significant difference in measured color between the out-of-eclipse and in-eclipse portions of the light curve \citep{wong2019}. The grazing nature of that shadowing event only allowed for a weak constraint on the spectral variability: a $1\sigma$ upper limit of 0.28\,mag on the difference in Sloan $g'-i'$ color between the two binary components, which translates to a $1\sigma$ upper limit of $\sim$10\slopeunit\ on spectral slope variations, assuming a linear continuum. Nevertheless, this constraint is sufficient to reject the scenario of a systematic intrabinary color difference at greater than $3\sigma$ confidence.

We also note that a large color difference between Patroclus and Menoetius contradicts the predictions from binary formation theories. While capture processes involving dynamical interactions \citep[e.g.,][]{goldreich2002} or coagulation of impact-generated debris \citep[e.g.,][]{durda2004} can yield spectrally heterogeneous pairs, the primary typically dominates the total system mass. Near-equal-mass binaries such as the PM binary, on the other hand, likely stem from coeval formation through the collapse of a single self-gravitating planetesimal cluster \citep[e.g.,][]{nesvorny2010}. This formation pathway is expected to produce companions with identical compositions and therefore similar surface properties.

%For a quantitative estimate of the colors of Patroclus and Menoetius in this scenario, we used the orbital solution of \citet{grundy2018} to calculate the expected occulted area of Menoetius during the 2024~Sep~2 observation: $A_{*} = 0.85A_{M}$, where $A_{M}$ is the sky-projected area of Menoetius. 

The second scenario assumes that both Patroclus and Menoetius have comparable overall spectral slopes, but there exist one or more isolated regions that have distinct properties from the rest of the surface regolith. In this instance, repeated observations of the PM binary at various rotational phases should retrieve a range of spectral slope values as the localized inhomogeneities rotate in and out of view. A recent work by \citet{fornasier2025} combined Gaia photometry from over a dozen individual visits and obtained a spectral slope value of $4.50\%\pm0.25$\slopeunit---intermediate between the two measurements reported from our GMOS observations targeting opposite hemispheric views.

Impact events provide a plausible avenue for imprinting small-scale spectral variability on Trojans. Photometric studies of the Eurybates and Ennomos collisional families have revealed that the fragments of Trojans are systematically bluer than the background population \citep[e.g.,][]{fornasier2007,deluise2010,wong2023}, indicating that the exposed surfaces resulting from impacts exhibit distinctive chemical and/or physical properties. As described in detail by \citet{wong2016}, this collisionally induced color shift is a natural outcome of Trojan formation and evolution models that posit an outer solar system origin \citep[e.g.,][]{morbidelli2005,levison2008}. The cold primordial surfaces of Trojans would have incorporated volatile ices such as methanol, which typically evolve into a darkened and reddened refractory crust upon irradiation \citep[e.g.,][]{brunetto2006}. 

Following the Trojans' inward migration to their present-day orbits, secondary thermal and/or physical processing of the surface could have removed or covered up this reddish outer layer. Given the significantly higher temperatures at $\sim$5\,au, any ice species exposed at the surface or situated in the near-surface interior, including water ice, would sublimate within yearlong timescales \citep[e.g.,][]{lisse2021}. Localized outgassing from subsurface layers can lead to the mechanical disruption of the refractory crust, as well as obscuration from the redeposition of rocky material entrained in the sublimating gas \citep[e.g.,][]{wong2024,fornasier2025}. Meanwhile, impacts readily destroy the outer irradiated layer and expose the Trojans' interior material, which subsequently undergoes rapid devolatilization. Without the retention of volatile ice species such as methanol or ethane that can effectuate surface reddening upon irradiation, the resultant surface would have a significantly more neutral color than the unaltered regolith. 

Within this explanatory framework, the low spectral slope value measured during the occultation of Menoetius on 2024~September~2 may be indicative of a region of impact-exposed subsurface material on the anti-Menoetius hemisphere of Patroclus. Alternatively, the spectrally anomalous region may lie within the unocculted portion of Menoetius. This latter possibility in particular is motivated by the report of a large, possibly impact-related topographic depression near Menoetius's south pole, which was detected from a superior occultation on UT 2017~December~8 \citep{pinillaalonso2022}. The viewing geometry during that observation (subobserver longitude and latitude of 180$^{\circ}$ and 7$^{\circ}$) was similar to that of our first GMOS visit. We recall that in our coordinate convention, the southern hemisphere of Menoetius is oriented toward positive declinations (i.e., upward in Figure~\ref{fig:geom}). It follows that the surface anomaly was at least partially in view during the 2024~September~2 GMOS visit and therefore may be responsible for the significantly shallower measured spectral slope. The true spatial extent of the topographic feature on Menoetius is poorly constrained, as it was only detected from a single occultation chord. High-precision measurements from future time-series observations of mutual events, combined with updated shape models for both binary components, will be needed to quantify the characteristics of the topographic depression on Menoetius.

%===============================================================================
\vspace{-0.7cm}

\begin{acknowledgments}

\facilities{Gemini South/GMOS-S.}
\software{\texttt{DRAGONS} \citep{dragons}, \texttt{emcee} \citep{emcee}, \texttt{matplotlib} \citep{matplotlib}, \texttt{numpy} \citep{numpy}, \texttt{scipy} \citep{scipy}}

\end{acknowledgments}

%===============================================================================
\vspace{-0.7cm}

\bibliography{main.bib}{}
\bibliographystyle{aasjournalv7}

%===============================================================================
\end{document}